\documentclass{iopjournal} 
\usepackage{float}
\usepackage{amsmath}
\usepackage{caption} 
\usepackage{braket}
\usepackage{cite}
\usepackage{marginnote} 

\begin{document}

\let\oldmarginpar\marginpar
\renewcommand{\marginpar}[1]{\marginnote{#1}}


\title{SFQED Cascade Saturation and Laser Depletion Dynamics In CP Rotating Laser Standing Waves}

\author{André Beverido$^*$\orcid{0000-0002-8023-2594}, Thomas Grismayer$^*$\orcid{0000-0002-0045-389X} and Marija Vranic$^{*}$\orcid{0000-0003-3764-0645}}

\affil{$^*$GoLP/Instituto de Plasmas e Fusao Nuclear, Instituto Superior Tecnico, Universidade de Lisboa, Lisboa 1049-001, Portugal} \\

\email{andre.beverido@tecnico.ulisboa.pt, thomas.grismayer@tecnico.ulisboa.pt, marija.vranic@tecnico.ulisboa.pt}

\begin{abstract}
Multi-petawatt laser facilities are approaching the regime where colliding laser pulses can spontaneously ignite self-sustained QED cascades, converting laser energy into dense electron-positron plasmas. Using 1D PIC-QED simulations of a circularly polarized standing wave, we characterize the complete lifecycle of seeded cascades from exponential growth through relativistic opacity and macroscopic laser depletion. We show that cascade saturation is governed by the competition between pair creation and longitudinal lepton migration out of the magnetic nodes. The overdense plasma reflects the driving laser, disrupting the standing wave and triggering secondary cascades in adjacent nodes, driving an apparent spatial expansion whose velocity scales linearly with the cascade growth rate. Finally, we generalize an analytical laser absorption model to realistic time-dependent pulse envelopes, accurately reproducing the simulated conversion efficiency of laser energy into gamma-ray photons and electron-positron pairs. 
\end{abstract}
\vspace{1.5cm} 

\let\marginpar\oldmarginpar


\section{Introduction}
The rapid development of multi-petawatt laser facilities~\cite{laser_power_in_time, eli, apolon, grafica_base_high_power_time, corels, xcels} promises access to electromagnetic field intensities exceeding $10^{23}\text{--}10^{24}\,\text{W/cm}^2$, opening an experimental regime where relativistic plasma dynamics is tightly coupled with strong-field quantum electrodynamics (SFQED)~\cite{DiPiazza2012, Gonoskov2022, Fedotov2023}, akin to extreme astrophysical environments such as pulsar magnetospheres~\cite{astro1, astro2}. In this ultra-intense regime, characterized by a normalized vector potential $a_0 = eE_0/(m_e c \omega_0) \gg 1$, electrons and positrons accelerated by the laser fields experience effective rest-frame fields that approach the Schwinger critical limit $E_S = m_e^2 c^3/(e\hbar)$~\cite{schwinger}. Under these conditions, particle motion is strongly affected by quantum radiation reaction~\cite{Neitz2013, Vranic_2016, Ridgers_2017, Niel2018} through nonlinear Compton scattering~\cite{klepikov, Nikishov1964, Ritus1985}, whereby leptons emit high-energy gamma-ray photons. As these emitted photons propagate through the intense background field, they can decay into electron-positron ($e^-e^+$) pairs via the multiphoton Breit-Wheeler process~\cite{klepikov, Nikishov1964, Ritus1985, SLAC}. Because the newly created pairs are themselves rapidly accelerated by the laser field and emit further hard photons, the interplay between these two processes can trigger an exponential, avalanche-type QED cascade \cite{Mironov-2014-PLA}, converting initially tenuous targets or seed electrons into a dense relativistic pair plasma.  Achieving a self-sustained QED cascade in the laboratory requires field configurations that maximize the quantum nonlinearity parameter $\chi_e$ of the interacting particles \cite{Gelfer2015, Vranic_2017, Gonoskov2017, Jirka2024}. A widely studied geometry is the circularly polarized (CP) standing wave formed by two counter-propagating laser pulses of equal wavelength and opposite helicity, originally proposed in this context by Bell and Kirk \cite{Bell2008} and examined in subsequent analytical and numerical studies. This setup establishes a periodic spatial structure where magnetic nodes ($B=0$) coincide with electric antinodes, producing a purely rotating electric field that continuously accelerates charges.

Zero-dimensional (0D) models restricted to a single magnetic node have served as the standard theoretical benchmark for estimating cascade growth rates and intensity thresholds \cite{Fedotov, Bashmakov2014, Grismayer2017}. However, as demonstrated by Kirk through a linear stability analysis of the classical phase-space trajectories \cite{Kirk_2016}, circular orbits located at the magnetic nodes of a CP standing wave are inherently unstable. Any longitudinal perturbation, whether arising from stochastic photon emission recoil or pair creation occurring slightly off-node, expels particles away from the $B=0$ regions. As these migrating leptons lose momentum through radiation damping, they become confined near the electric nodes ($E=0$), which act as phase-space attractors in the regime of normal radiative trapping (NRT) \cite{gonoskov2014anomalous, esirkepov2015attractors, Kirk_2016}. More recently, Mercuri-Baron et al. \cite{Mercuri2025} incorporated this longitudinal lepton migration into a one-dimensional (1D) theoretical framework, showing that the escape of particles from the magnetic nodes at a characteristic rate $\nu_e \sim \omega_0 / 2\pi$ significantly lessens the cascade growth near the pair-creation threshold. While the onset and exponential growth rate $\Gamma$ of cascades in standing waves are better understood in the test-particle limit, the macroscopic evolution of the system once the generated plasma becomes dense enough to modify the driving fields remains a central problem for upcoming experiments. If pair creation at the magnetic node overcomes longitudinal migration ($\Gamma > \nu_e$), the pair density grows exponentially until the plasma frequency becomes comparable to the relativistically corrected laser frequency, reaching the relativistic critical density $n_p \sim a_0 n_c$. At this stage, collective plasma currents screen and disrupt the standing wave, terminating the ideal rotating-field configuration and depleting the laser pulse energy \cite{Nerush2011, Bradly2012, Grismayer2016}. Understanding how longitudinal migration competes with pair multiplication to set the threshold for relativistic opacity and how the standing-wave breakdown alters subsequent pair production, plasma expansion, and laser absorption is essential to identify robust macroscopic diagnostics of QED cascades in multi-petawatt laser interactions. 

In this paper, we investigate the spatiotemporal lifecycle of seeded QED cascades in a $1 \mathrm{\mu m}$ laser rotating CP standing wave using self-consistent 1D3V particle-in-cell (PIC) QED simulations with the code OSIRIS \cite{OSIRIS, osiris_qed, marija_phd, Vranic_2016}. By performing a parametric study over single-laser amplitudes ranging from $a_0 = 300$ to $a_0 = 1000$, we bridge the gap between early migration-limited cascade growth and late-time nonlinear saturation. We show that the competition between the cascade growth rate and the lepton migration rate produces two different regimes. The migration-dominated regime $\Gamma < \nu_{e}$ is where particles diffuse too fast from the particle creation zone, lessening the cascade development, and the strong particle creation regime, where the peak density reaches the relativistic critical density. Furthermore, we analyze the disruption of the standing wave via the electric energy density and Poynting flux, showing that wave reflection at the critical surface creates a dynamic interference pattern that shifts the phase-space attractors and triggers secondary cascades in adjacent nodes. This process drives an apparent spatial expansion of the over-dense plasma whose velocity scales linearly with the cascade growth rate. Finally, building on the analytical absorption model \cite{Grismayer2016}, we introduce an adiabatic generalization for arbitrary temporal laser envelopes that accurately predicts the macroscopic laser energy depletion across the high-intensity regime.  

This paper is organized as follows. Section 2 reviews the theoretical background of SFQED processes, the field structure and radiative trapping dynamics in a CP standing wave, and the dispersion relation governing relativistic opacity in a pair plasma.
Section 3 describes the 1D3V PIC-QED simulation setup and numerical parameters. In Section 4, we present our numerical results on the cascade growth rate across the migration and strong-creation regimes, the spatiotemporal confinement of the plasma, the density overshoot at relativistic opacity, and the apparent expansion driven by standing-wave disruption. Section 5 examines the partitioning of the absorbed laser energy and presents our extended analytical model for laser energy depletion in realistic pulse envelopes. Finally, Section 6 summarizes our main conclusions and discusses their implications for experimental diagnostics at upcoming multi-petawatt laser facilities.

\section{Theoretical Background}
A particle can experience quantum strong-field phenomena if the field it encounters approaches the Schwinger limit \cite{schwinger} and $a_{0} \gg 1$. Although the laser fields are many orders of magnitude below the Schwinger critical field in the laboratory, a relativistic particle can boost the effective field it experiences in its rest frame to values approaching the Schwinger field. The strength of this effect is captured by the quantum nonlinearity parameter
\begin{equation}
    \chi_e = \gamma \frac{\sqrt{(\vec{E}+\vec{v}\times \vec{B})^{2} -(\vec{E}\cdot\vec{v}/c)^{2}}}{E_{S}}.
\end{equation}
When $\chi_e$ approaches unity, two processes become dominant. In nonlinear Compton scattering, a relativistic electron or positron absorbs multiple laser photons and emits a single high-energy gamma-ray photon. In the nonlinear Breit-Wheeler process, such an emitted gamma-ray photon propagating through the intense background field decays into an electron-positron pair. Because the newly created pairs are themselves rapidly accelerated by the laser and emit further hard photons, the interplay between these two processes can trigger an exponential, avalanche-type QED cascade \cite{Nerush2011, Mironov-2014-PLA, Bashmakov2014, Grismayer2017, Mercuri2025}, converting an initially tenuous seed plasma into a dense relativistic pair plasma.

\subsection{CP Standing Wave}

The CP standing wave, originally proposed by Bell and Kirk \cite{Bell2008} and extensively studied since \cite{Elkina2011, Nerush2011, Bashmakov2014, Grismayer2016}, maximizes $\chi_e$ at the magnetic nodes where the two counter-propagating fields interfere constructively, producing a purely rotating electric field. Figure (\ref{fig:laser configuration diagram}) shows a schematic of this configuration, and the individual laser fields can be represented as:
\begin{figure}[H]
    \centering
    \includegraphics[width=0.85\textwidth]{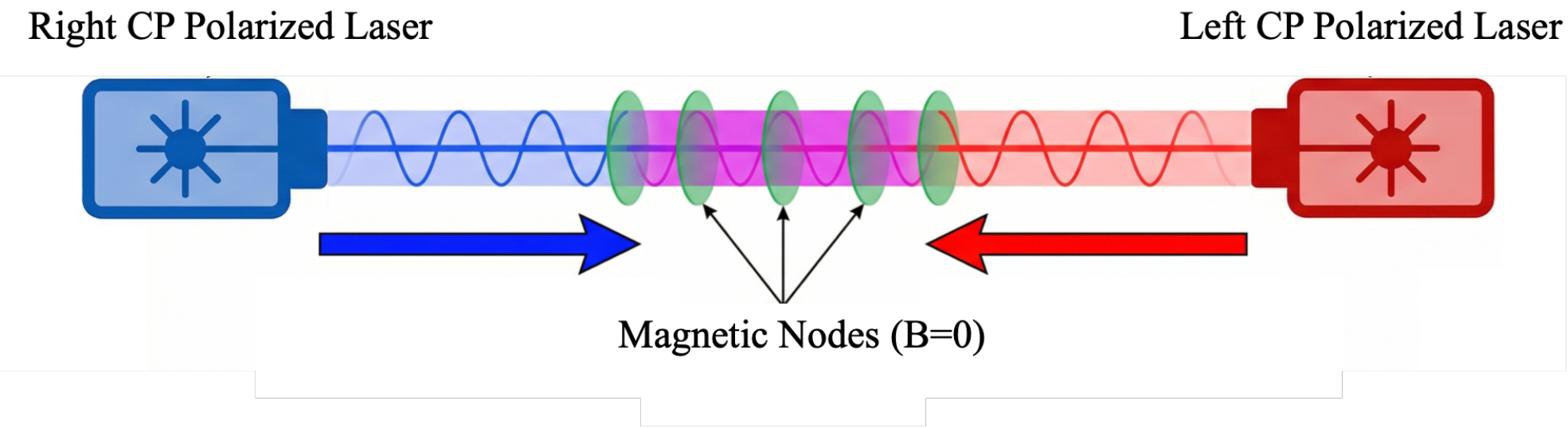}
    \caption{Schematic of the CP standing wave configuration. Two circularly polarized counter-propagating lasers with equal wavelengths and different helicity create a standing wave with defined electric and magnetic nodes. The colors are meant to differentiate the lasers in the schematic; both have the same wavelength.}
    \label{fig:laser configuration diagram}
\end{figure}
The laser electric fields are
\begin{align*}
    & \vec{E}_{1}(x,t) = E_{0} \left[ \cos(kx - \omega t)\hat{{y}} +\sin(kx-\omega t) \hat{z}\right] \\
    & \vec{E}_{2}(x,t) = E_{0} \left[ \cos(kx + \omega t)\hat{{y}} -\sin(kx+\omega t) \hat{z}\right].
\end{align*}
The total electric and magnetic fields that describe the standing wave are given by
\begin{align}
    & \vec{E}_{tot}(x,t) = 2E_{0}\cos(kx)\left[\cos(\omega t)\hat{y} - \sin(\omega t)\hat{z} \right],\\
    & \vec{B}_{tot}(x,t) = 2(E_{0}/c)\sin(kx)[-\cos(\omega t)\hat{y} + \sin(\omega t) \hat{z} ].
\end{align}
where $\nabla \times \vec{E}_{tot} = - \partial_{t}\vec{B}_{tot}$. Figure (\ref{fig: standing wave diagram}) shows a schematic of the formed standing wave. The linear stability analysis of the phase-space Jacobian by Kirk \cite{Kirk_2016} showed that circular orbits located at the magnetic nodes correspond to unstable trajectories. Any perturbation causes particles to migrate from the magnetic nodes and become trapped in the vicinity of the electric nodes, which act as phase-space attractors where particles accumulate. This phenomenon has been labeled as Normal Radiative Trapping (NRT) corresponding to the laser amplitude regime of $1 \ll a_{0} < 10^{4}$ \cite{gonoskov2014anomalous, esirkepov2015attractors, Kirk_2016}.
\begin{figure}[H]
    \centering
    \includegraphics[width=0.85\textwidth]{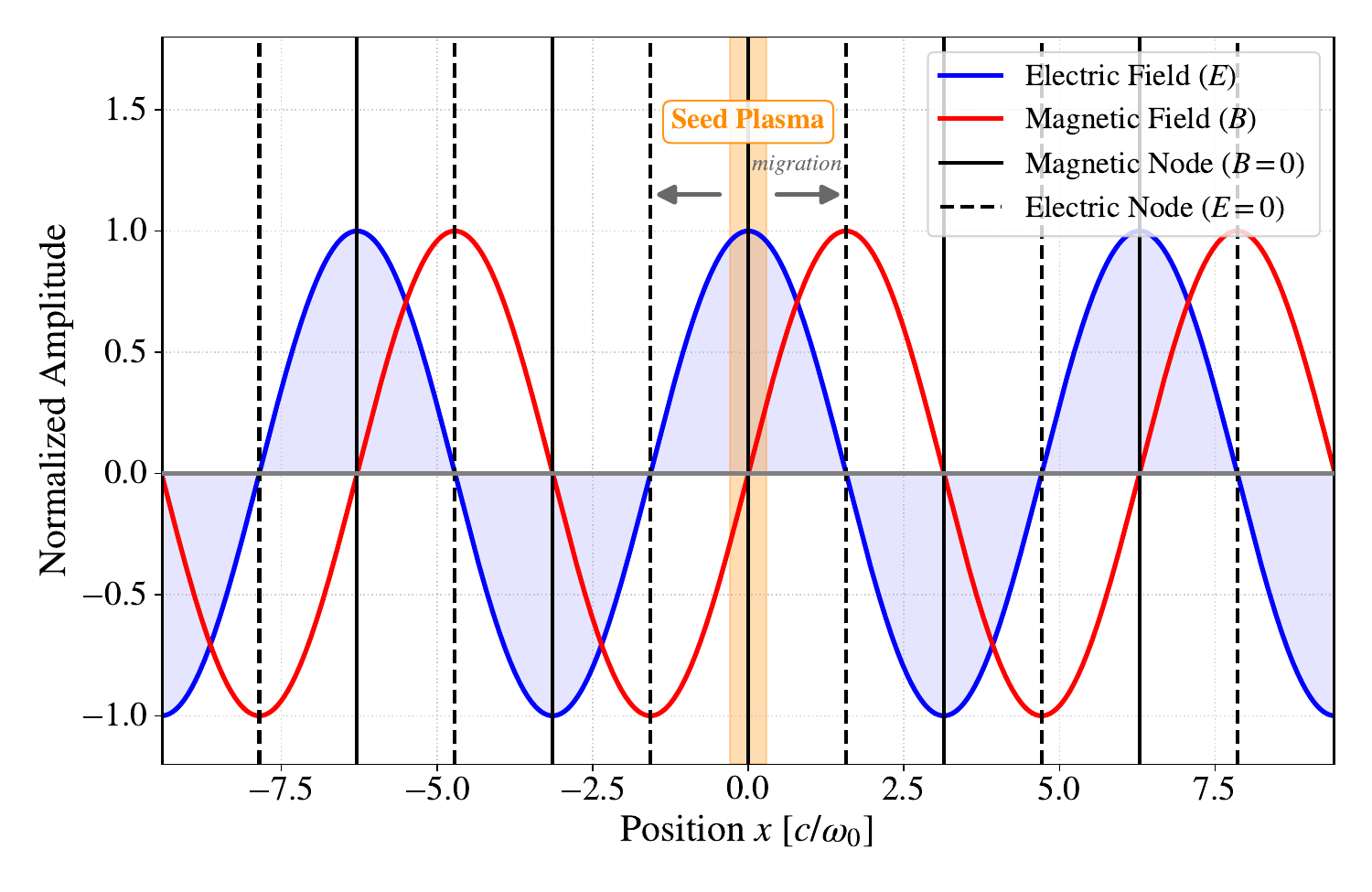}
    \caption{Standing wave schematic for the CP configuration. The configuration shows the interplay between the electric and magnetic nodes in space.}
    \label{fig: standing wave diagram}
\end{figure}
For a pair plasma interacting with an ultra-intense wave $a_0 \gg 1$, nonlinearities modify the dispersion relation obtained for a proton-electron plasma \cite{Iwamoto2026}, raising the critical density at which the plasma becomes opaque to the laser to $n_p = a_0 n_c$. Once the cascade-generated plasma reaches this relativistic critical density, collective plasma currents screen the driving fields and the standing wave configuration is disrupted. The conditions under which this threshold is reached, and what follows from it, are the subject of Section 4.

\section{Simulation Setup}
All simulations are performed with the fully relativistic 1D3V particle-in-cell code OSIRIS \cite{OSIRIS, osiris_qed} , extended to include strong-field QED effects: photon emission via nonlinear Compton scattering, classical and quantum radiation reaction damping, and electron-positron pair production via the nonlinear Breit-Wheeler process.

To model the standing-wave configuration, we use a simulation box of length $L = 41\lambda_0$, where $\lambda_0 \simeq 1\,\mu\mathrm{m}$ is the laser wavelength and $\omega_{p0}$ is the reference frequency ($\omega_0/\omega_{p0} = 1$). 
The box is initialized such that it is centered at a magnetic node ($B=0$). 
The two lasers were initialized by two collinear counter-propagating plane waves with circular polarization (right-handed and left-handed, respectively). The temporal envelope of the lasers is composed by a $\sin^{2}$ rise, followed by a region of constant amplitude $a_{0}$, and a $\sin^{2}$ fall. This envelope was chosen as a physically realistic alternative to an ideal flat-top pulse, allowing us to capture the cascade dynamics during both the intensity ramp and the flat-top phase, while remaining comparable to the flat-top theoretical predictions of \cite{Grismayer2016}. Both laser pulses in the simulation have a duration of 13 periods ($T \approx 3.3 \mathrm{fs} $)  which corresponds to a  $42.5\ \mathrm{fs}$ duration distributed in $11\ \mathrm{fs}$ distributed evenly between the laser rise and fall and $31.5 \ \mathrm{fs}$ of constant amplitude.

The spatial resolution is set to $\Delta x \approx \lambda_0 / 120$. To accurately resolve the particle dynamics in the ultra-relativistic regime, the temporal resolution is strictly chosen as $\Delta t = T / 12566$. This choice safely satisfies the convergence condition $\Delta t \ll \min(T/500, T/a_0)$ 
for our maximum explored single-laser amplitude of $a_0 = 1000$. 
At the center of the domain ($x = 128\, c/\omega_{p0}$), we initialize a pre-formed electron-positron seed plasma uniformly distributed within a narrow slab of thickness $d = 0.2\, c/\omega_{p0} \approx 31.8\,\mathrm{nm}$, with an initial density of $n_{e^-} = n_{e^+} = 0.1\ n_c$ for each species, where $ n_c$ is the classical critical density. The plasma is modeled using 10 macroparticles per cell for both species, with zero initial thermal momentum. Simulations initialized with a purely electron seed plasma were also performed and produced identical cascade dynamics, confirming that the results are insensitive to the initial plasma composition.

\section{Cascade Development}

\subsection{The QED Cascade Growth Threshold}

To systematically characterize cascade formation across the intensity range accessible to upcoming multi-petawatt facilities, we performed a parametric study over single-laser amplitudes $a_{0} \in [300,1000]$. For each amplitude, the cascade growth rate was extracted from the simulated particle count during the steady-state pair creation phase using a logarithmic fit. These measured growth rates are compared in Figure 3 against the models by Mercuri-Baron et al. \cite{Mercuri2025}: the 0D model, which neglects all particle migration, and the 1D model, which accounts for lepton migration while neglecting photon migration. 
Previous studies have utilized a 0D model based on a rotating electric field configuration \cite{Nerush2011, Elkina2011,Fedotov,Grismayer2017} when $\chi_e \ll 1$ and $\chi_e \gg 1$. This model covers the regime $\chi_e \sim 1$ and predicts growth rates 
\begin{equation}
    \frac{\Gamma}{\omega_{0}} = \frac{W_{cr} + \nu_{e} + \nu_{\gamma}}{2} \left[-1 + \sqrt{1 + 4\frac{W_{cr}(2W_{rad}-\nu_{e})-\nu_{e}\nu_{\gamma}}{(W_{cr}+\nu_{e}+\nu_{\gamma})^{2}}} \right],
\end{equation}
with the two asymptotic regimes
\begin{align}
\frac{\Gamma}{\omega_{0}} &= 1.23\sqrt{\frac{\alpha a_{0}}{2}}\exp\left(-0.96\frac{\alpha a_{\mathrm{S}}}{a_{0}^{\mathrm{eff}}}\right)~~~(\nu = 0) \\
\frac{\Gamma}{\omega_0} &= 1.52 \alpha a_0 \left( \omega_{0}/2\nu_e \right) \exp\left(-1.92 \frac{\alpha a_{\mathrm{S}}}{a_0^{\mathrm{eff}}}\right)~~~(\Gamma < \nu) 
\end{align}
$W_{rad}, W_{cr}$ are the effective rates for photon emission and pair creation, respectively; $a_{0}^{\mathrm{eff}}=2a_{0}$, $\alpha$ is the fine structure constant, $a_{\mathrm{S}}$ is the normalized Schwinger field amplitude, and $\nu_{e} = \omega_{0}/2\pi$ is the fiducial lepton migration rate.

The simulation data reveal three distinct regimes as the laser amplitude increases, illustrated in Figure 3. At low amplitudes $300 \le a_{0} \le 400$, the system operates in the migration-dominated regime $\Gamma \ll \nu_{e}$: pair creation occurs, but lepton migration depletes the magnetic node faster than pairs accumulate, suppressing cascade development and confining the plasma to low densities. In the intermediate range $400 \le a_{0} \le 600$, the system enters a transition regime where pair creation and migration are comparable $\Gamma \approx \nu_{e}$: the cascade is established, but migration still partially suppresses its growth, and the simulation data follow the 1D model closely. Above $a_0 = 600$, the system crosses into the strong creation regime ($\Gamma > \nu_e$), where pairs are created faster than they migrate, and the data converge toward the 0D model, consistent with the prediction \cite{Mercuri2025} that the 1D model collapses into the 0D model as the field strength increases.

\begin{figure}[H]
\centering\includegraphics[width=0.85\textwidth]{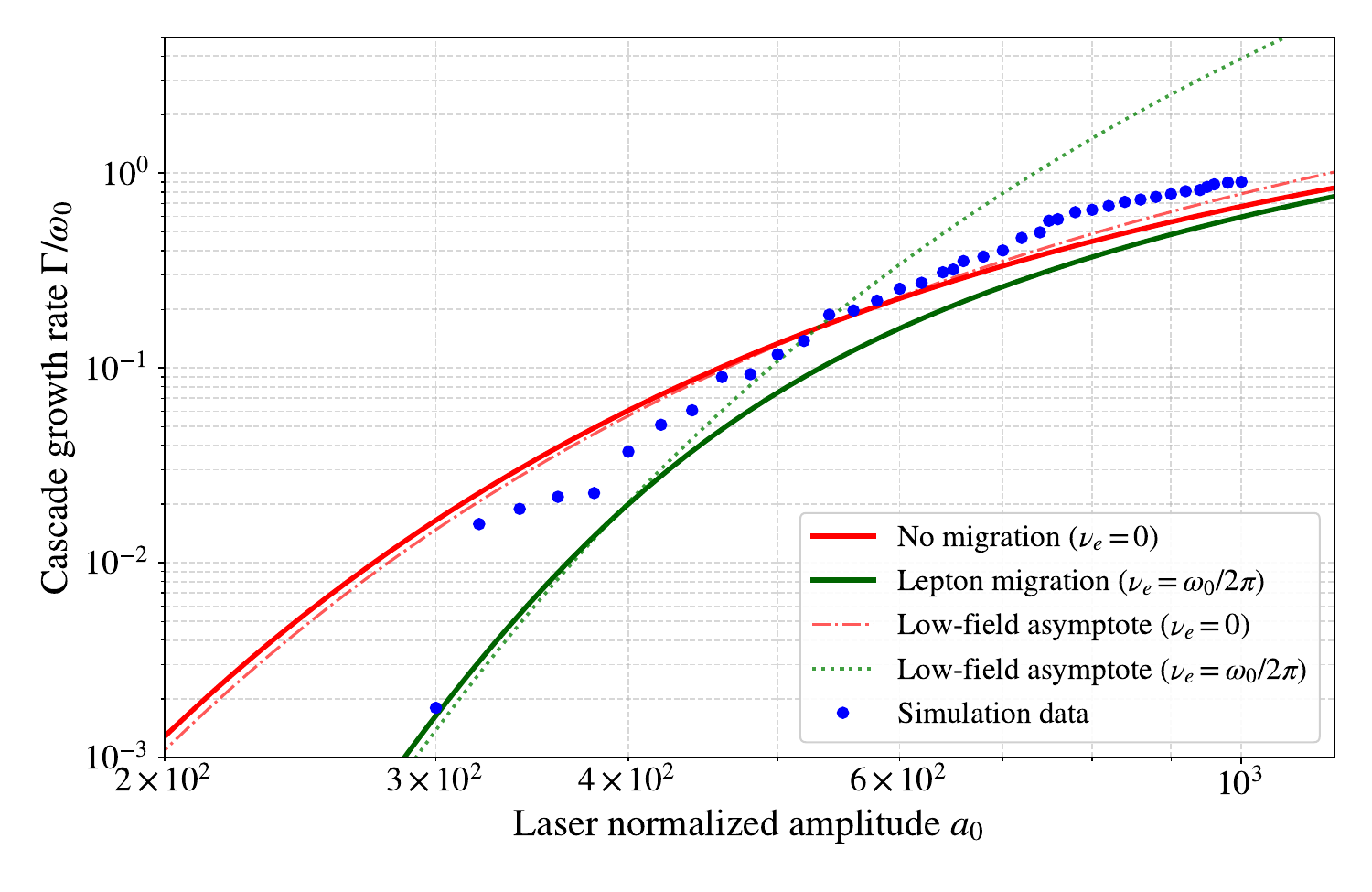}
\caption{Growth rate for a parameter scan in the CP standing wave configuration. The data are compared with the theoretical asymptotic behavior of the 0D model without migration (red curve) and the 1D model with low-field lepton migration (green curve) derived in \cite{Mercuri2025}. At low fields, the data follow the migration curve trend, but as the field strength increases, they transition to a stronger particle creation regime.}
\label{fig:Growth rate}
\end{figure}

\begin{figure}
\centering\includegraphics[width=0.85\textwidth]{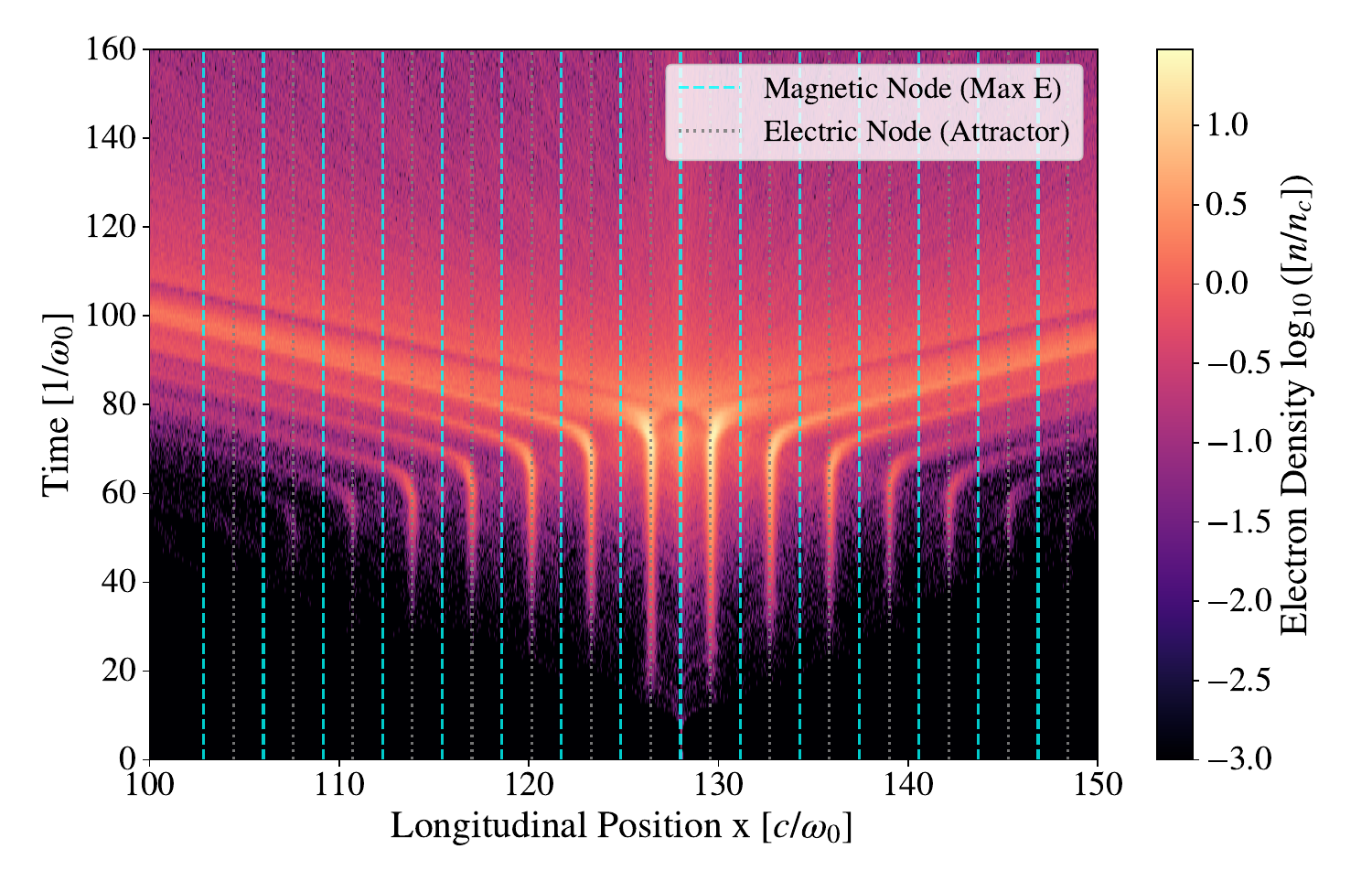}
\caption{Spatiotemporal evolution of the plasma electron density formed by a QED cascade in the CP rotating standing wave configuration for a single-laser amplitude of $a_0 = 500$. The waterfall plot depicts the laser interacting with the seed plasma to generate a QED cascade, which ultimately forms a dense plasma. Particle migration is evident, as regions of high density correspond to the positions of the classical normal radiative trapping attractors \cite{gonoskov2014anomalous, esirkepov2015attractors, Kirk_2016}.}
\label{fig:plasma waterfall 500 lines}
\end{figure}

These threshold amplitudes are wavelength-dependent, since the quantum parameter at the magnetic node scales as $\chi_e = a_0^2/a_s \propto \lambda$. For the laser wavelength $\lambda = 1 \mathrm{\mu m}$ adopted here, the results are directly relevant to current and upcoming multi-petawatt facilities, most of which operate near this wavelength. With the growth rate behaviour now established across the full parameter scan, we turn to the spatial development of the cascade and the conditions under which a dense, confined pair plasma is formed.

\subsection{Particle Confinement and Relativistic Opacity }
Having established the cascade growth rates across the parameter scan, we now turn to the spatial development of the plasma and the conditions under which it reaches relativistic opacity. 

Figure \ref{fig:plasma waterfall 500 lines} illustrates the spatiotemporal formation of the plasma for $a_0 = 500$ via a waterfall plot of the electron density. The onset of the cascade is clearly visible, with the highest-density regions coinciding with the electric nodes, where Normal Radiative Trapping confines the migrating particles \cite{Kirk_2016}. The resulting density profile exhibits the standing-wave-like pattern characteristic of the CP configuration, with particle migration progressively populating electric node attractors across the full simulation box.

\begin{figure}
\centering\includegraphics[width=\textwidth]{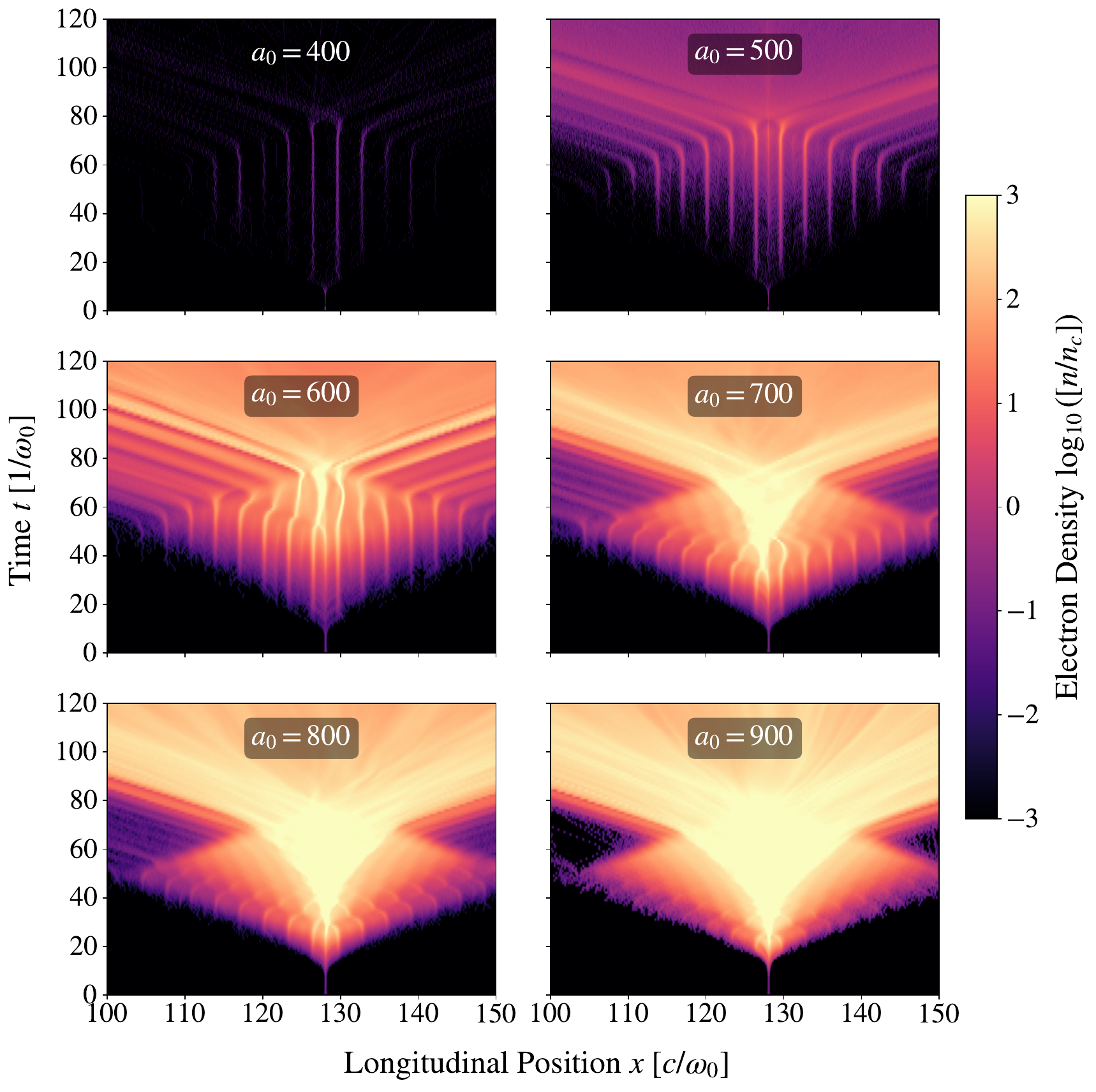}
\caption{Spatiotemporal evolution of the pair-plasma electron density generated by the QED cascade for various single-laser amplitudes $a_{0}$ in the rotating field configuration. The plasma reaches progressively higher densities at the principal node as the field strength increases.}
\label{fig:waterfall_a0_scan}
\end{figure}

Figure \ref{fig:waterfall_a0_scan} extends this picture across the full parameter scan. In the migration-dominated regime $\Gamma < \nu_{e}$ ($a_{0} \in [400,500]$), particles diffuse too rapidly out of the magnetic nodes to sustain high local densities, and the plasma retains a well-defined standing-wave pattern at all times. As the system transitions into the strong creation regime $\Gamma > \nu_{e}$ ($a_{0} > 600$), pair creation becomes prolific enough to overcome migration losses. The standing-wave density pattern begins to dissolve, reaching values that exceed the relativistic critical density not only at the central node but across all attractors. For amplitudes  $a_{0} > 700$, a characteristic V-shaped interface surface emerges in the waterfall plots, becoming increasingly steep as $a_0$ rises. As we demonstrate below, this conical surface corresponds to the relativistic critical density boundary beyond which the laser can no longer interact with the plasma.

\begin{figure}[H]
\centering\includegraphics[width=0.85\textwidth]{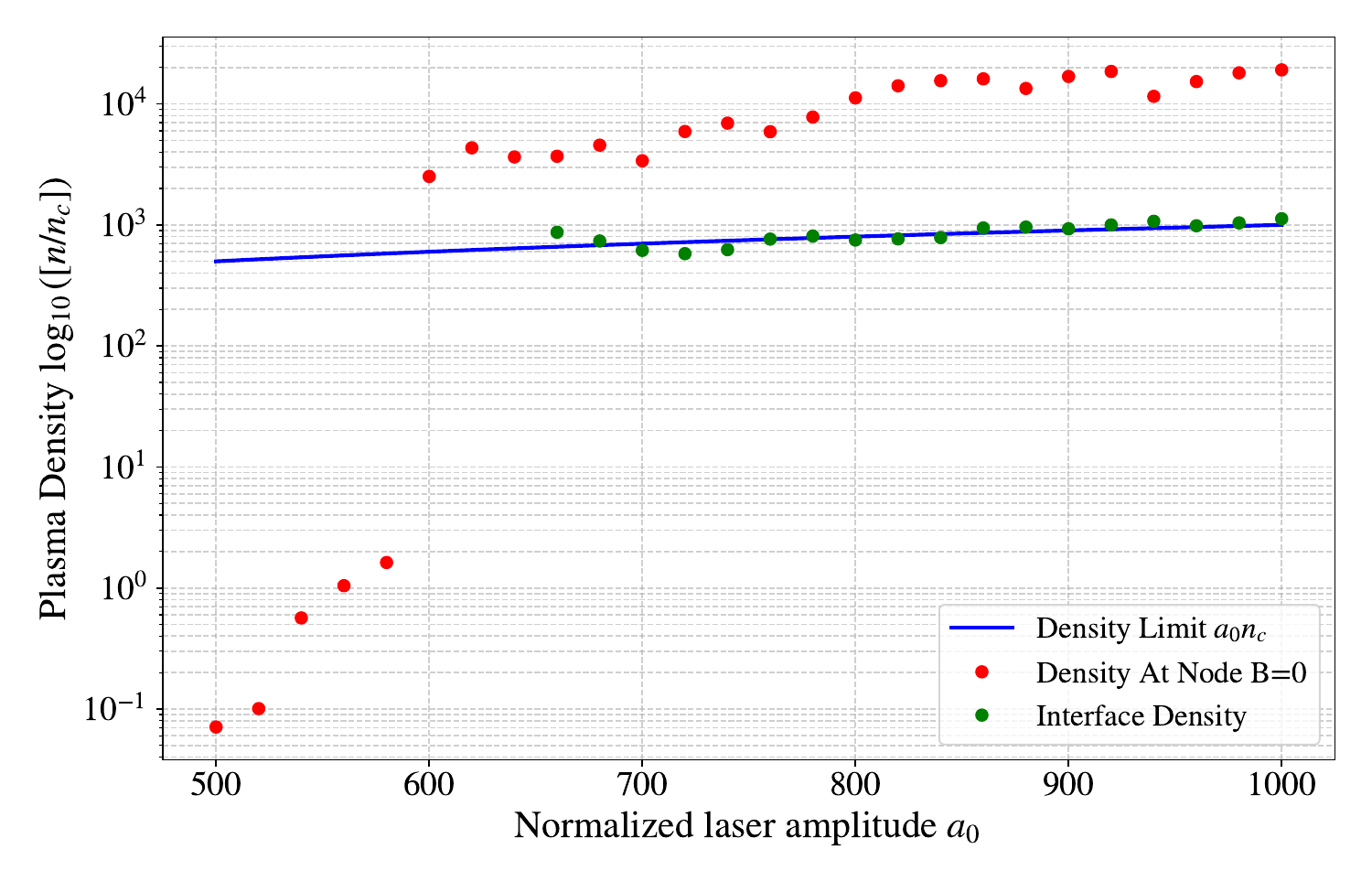}
\caption{Plasma density generated by the QED cascade in the CP standing wave configuration. The red markers indicate the density overshoot values at the magnetic node, while the green markers represent the density at the V-shaped relativistic critical density surface. The solid line denotes the relativistic critical density at the magnetic node, $a_{0}n_{c}$. The data show a density jump across laser amplitudes, dictated by the interplay between particle migration ($\nu_{e}$) and particle creation strength ($\Gamma$). Driven by particle creation inertia, the cascade produces an overdense pair plasma at the central magnetic node for single-laser amplitudes $a_{0} \ge 600$.}
\label{fig:max density}
\end{figure}

Figure \ref{fig:max density} quantifies this behaviour by tracking the peak plasma density at the central magnetic node and at the V-shaped interface across the parameter scan. The data reveal a sharp density jump across the amplitude range $500 \le a_{0} \le 600$, marking the transition between the two dynamical regimes identified in Section 4.1. Below this threshold, the plasma remains underdense as migration continuously suppresses cascade development ($\Gamma < \nu_{e}$). Above it, the cascade drives the node density well beyond the relativistic critical density and, in some cases, by nearly an order of magnitude. This pronounced density overshoot is a direct consequence of what we term particle creation inertia. As the plasma approaches the relativistic critical density, it ceases to interact with the standing wave, but cascade development does not halt abruptly. Particles retain high energies and continue emitting hard photons, while the background laser field is simultaneously being absorbed. These accumulated high-energy photons continue producing pairs even as the driving field weakens, driving the node density significantly above the opacity threshold before the remaining laser energy is fully depleted.

To confirm that the cascade-generated plasma constitutes a genuine plasma rather than a mere dense particle distribution, we verify the relevant plasma formation criteria at late times ($t \simeq 80$) for $a_0 > 700$. The particle density reaches $n \sim 100\ n_c$ over a longitudinal extent $L \ge 2.4 \, \mathrm{\mu m}$. The Debye length in the relativistic regime satisfies $\lambda_{D} \sim (c/\omega_{p})\sqrt{\gamma}$, which for $\lambda = 1 \, \mathrm{\mu m}$ yields $n\lambda_{D}^{3} \sim 10^{5} \gg 1$ and $L > \lambda_{D} \sim 0.63 \, \mathrm{\mu m}$, explicitly satisfying the fundamental criteria for plasma formation. Having confirmed that an overdense pair plasma is established, we now examine its consequences for the driving laser fields and the subsequent dynamics of the standing wave.

\subsection{Field Disruption and Plasma Expansion}
Once the cascade-generated plasma exceeds the relativistic critical density, the incident laser pulses can no longer penetrate it and are instead reflected at the plasma boundary. This reflection disrupts the standing wave, replacing the previously stable periodic node structure with a time-dependent interference pattern between the incoming and reflected pulses. As we now show, this disruption does not simply terminate the cascade; it actively drives its spatial expansion into adjacent nodes.

\begin{figure}[H]
\centering\includegraphics[width=0.85\textwidth]{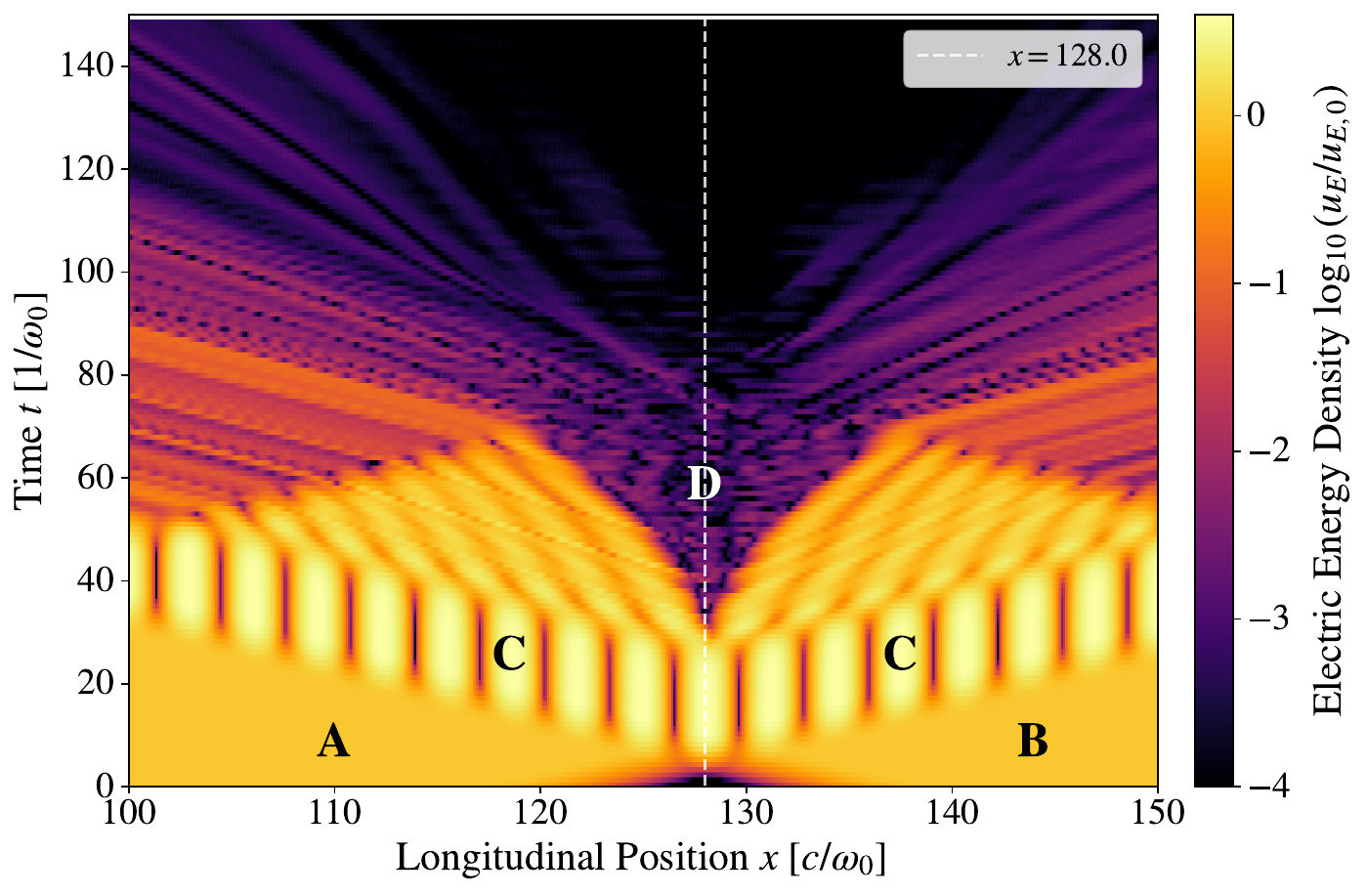}
\caption{Electric energy density of a self-sustained QED cascade in the CP standing wave configuration for a laser amplitude of $a_0 = 800$. The plot displays the central magnetic node at $x = 128$ alongside the disruption of the standing wave. The electric-field nodes and antinodes remain distinct until the critical plasma density is reached, after which a new, dynamic interference pattern forms. The regions are defined as follows: Region A corresponds to the right-propagating laser, Region B to the left-propagating laser, Region C to the standing-wave region, and Region D to the plasma-expansion region.}
\label{fig:E energy waterfall}
\end{figure}

Figure \ref{fig:E energy waterfall} illustrates this process through the spatiotemporal evolution of the electric energy density for $a_0 = 800$. Four distinct regions can be identified. Regions A and B correspond to the incoming right- and left-propagating laser pulses, where the electric energy density is maximal. Region C represents the standing wave before the plasma formation, characterized by the well-defined alternating pattern of electric energy maxima and minima. Region D marks the onset of plasma-induced disruption. Once the central node reaches the relativistic critical density, the standing wave collapses and is replaced by a dynamic interference pattern whose node positions shift continuously in time. This shifting pattern maintains a well-defined electric and magnetic node structure, enabling trapped particles to re-enter the particle creation zones and sustain further cascade activity.

\begin{figure}[H]
\centering\includegraphics[width=0.85\textwidth]{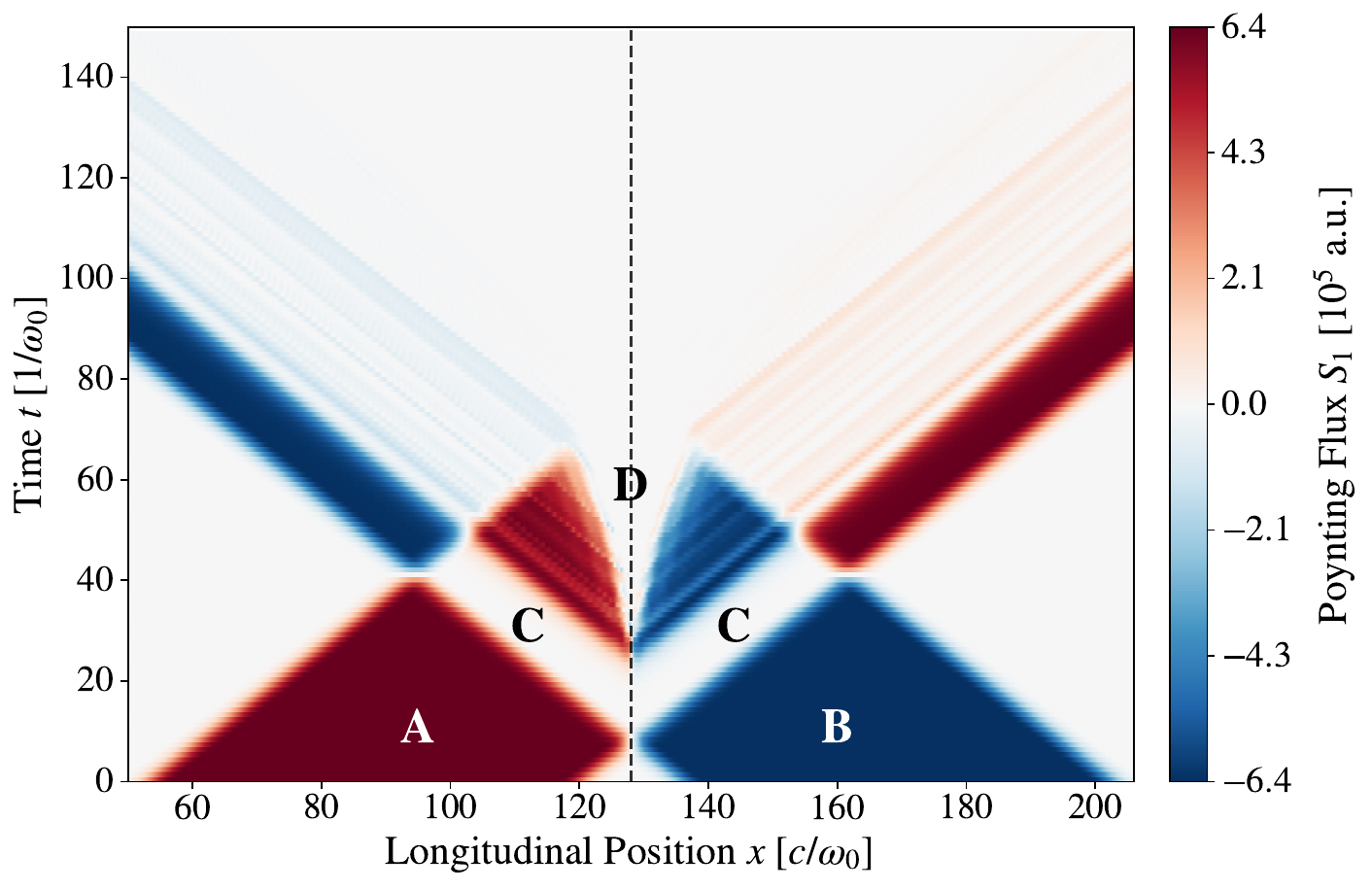}
\caption{Poynting flux of a self-sustained QED cascade in the CP standing wave configuration for a laser amplitude of $a_0 = 800$. The plot illustrates the spatiotemporal evolution of the longitudinal Poynting flux. Region A corresponds to the right-propagating laser, Region B to the left-propagating laser, Region C to the standing-wave region, and Region D to the plasma-expansion region. Although Regions C and D both exhibit zero net flux, this behavior stems from distinct physical mechanisms. At the plasma-expansion boundary, the Poynting flux changes sign, indicating wave reflection.}
\label{fig:Poynting waterfall}
\end{figure}

Complementary evidence for laser reflection is provided by the spatiotemporal evolution of the Poynting flux \cite{landau, jackson}, shown in Figure \ref{fig:Poynting waterfall} for the same amplitude. In Region C, the Poynting flux vanishes due to the superposition of two equal counter-propagating fluxes (the hallmark of a standing wave). In Region D, the flux also vanishes on average, but for a physically distinct reason: the overdense plasma excludes electromagnetic energy from its interior. At the boundaries of Region D, the Poynting flux changes sign, providing direct evidence of wave reflection. A fraction of the laser energy inevitably escapes absorption at the edges of this region, establishing an upper bound on the conversion efficiency.

\begin{figure}[H]
\centering\includegraphics[width=0.85\textwidth]{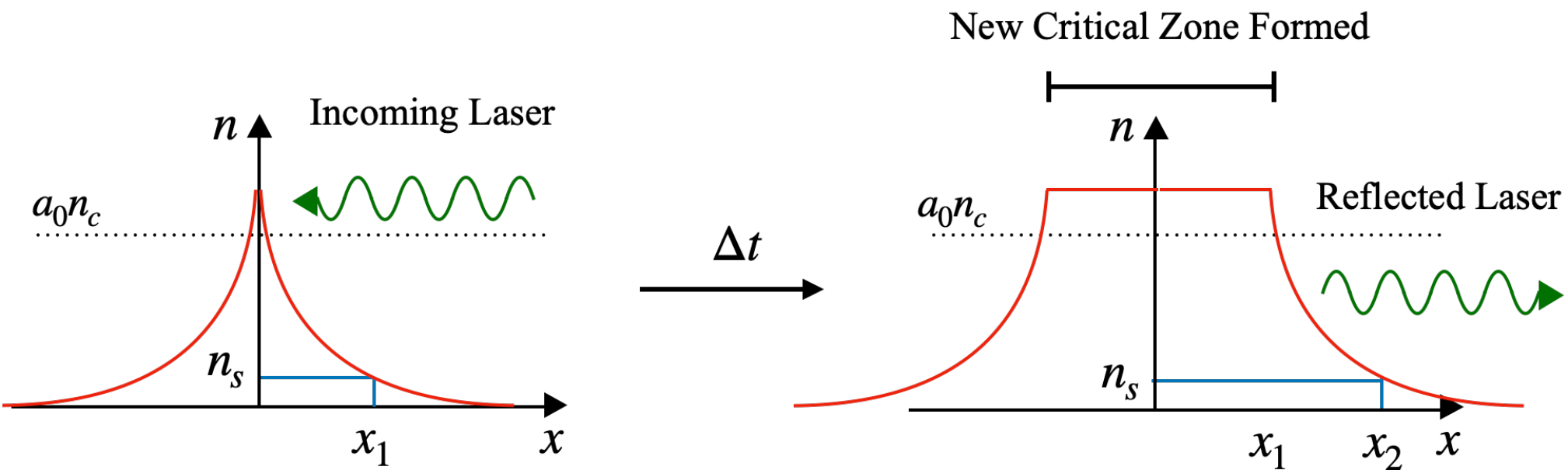}
\caption{Schematic of the apparent plasma expansion process. At a first time, the plasma generated by the initial QED cascade reaches the relativistic critical density, disrupting the standing wave. After an amount of time $\Delta t$, the disruption triggers secondary QED cascades that produce an overdense plasma ($n > a_{0}n_{c}$), resulting in the apparent spatial expansion.}
\label{fig:density creation schematic}
\end{figure}

The physical mechanism driving the apparent spatial expansion of the plasma is illustrated schematically in Figure \ref{fig:density creation schematic}. When the plasma at the central magnetic node reaches the relativistic critical density, it reflects the incoming laser and generates a new local standing wave whose node positions are shifted relative to the original configuration. Particles previously trapped at the electric nodes of the original standing wave re-enter the magnetic nodes of this new local wave, triggering secondary QED cascades. These secondary cascades drive adjacent nodes to the relativistic critical density, producing an expansion-like progression of the critical density surface outward from the initial seed position.

\begin{figure}
\centering\includegraphics[width=0.85\textwidth]{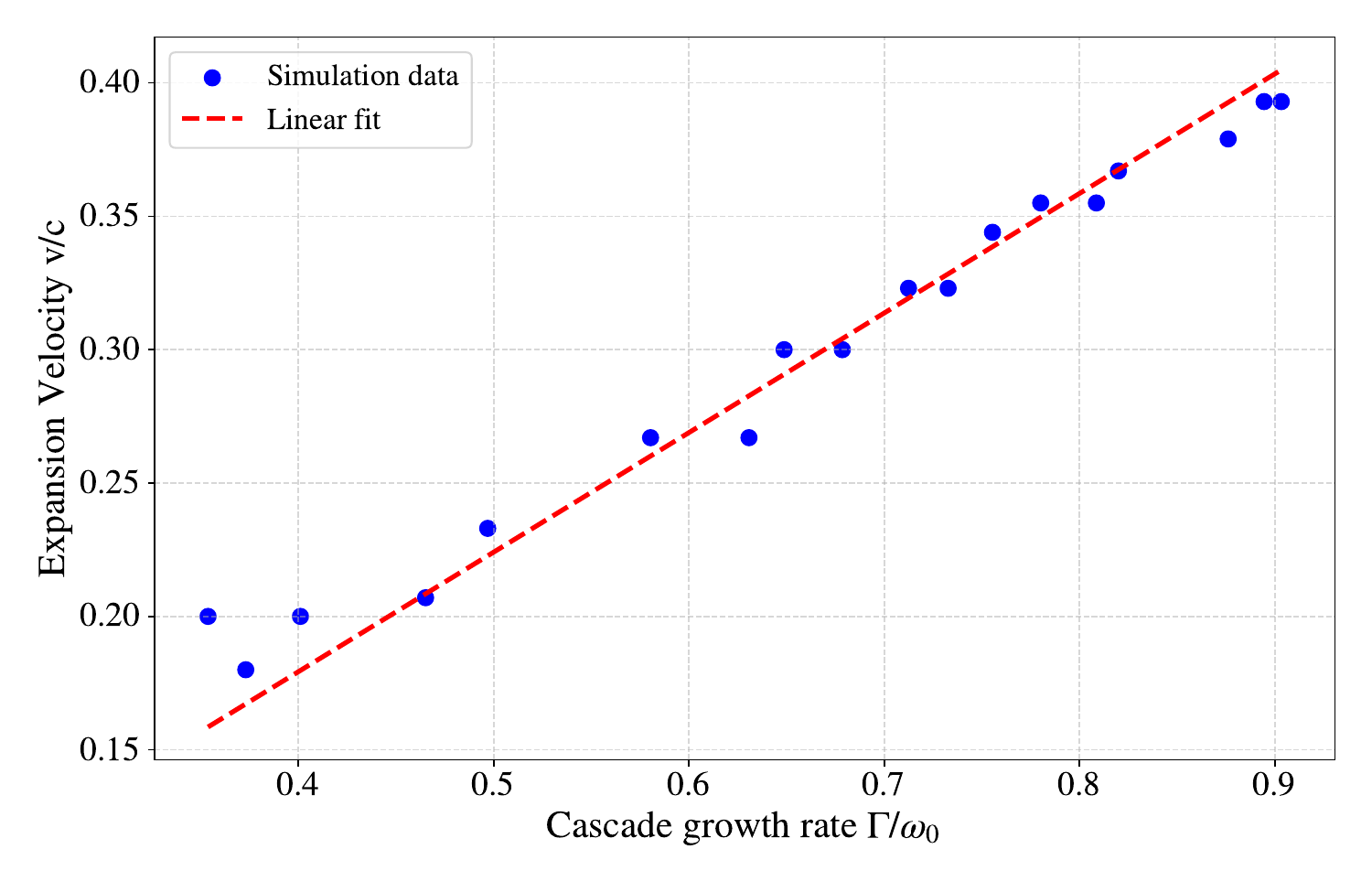}
\caption{Expansion velocity of the plasma generated by a QED cascade in the CP standing wave configuration. A linear fit describes the expansion according to the relation $v/c = 0.45 \frac{\Gamma}{\omega_0}$.}
\label{fig:Plasma expansion}
\end{figure}

The velocity of this apparent expansion can be estimated by noting that each successive node, separated by $\lambda/4$, reaches the critical density after an absorption time $t_{a} = (1/\Gamma)ln(a_{0}n_{c}/n_{s})$ \cite{Grismayer2016}. Assuming a constant laser pump, the expansion velocity is
\begin{equation}\label{expansion velocity expression}
    \frac{v}{c} \propto \frac{\Delta x}{c \Delta t} = \frac{\lambda \ \Gamma}{4c \ \ln(a_{0}n_{c}/n_{s})} = \frac{\pi}{2\ \ln(a_{0}n_{c}/n_{s})}\frac{\Gamma}{\omega_{0}}
\end{equation}
The logarithmic factor can be bounded by noting that $a_{0}n_{c} > n_{s} > n_{0}$, with $n_{0}=0.1n_{c}$, which constrains $\ln(a_{0}n_{c}/n_{s})$ to the range $2 < \ln(a_{0}n_{c}/n_{s}) < 8$ across our parameter scan. Figure \ref{fig:Plasma expansion} presents the expansion velocities measured from the waterfall plots of Fig. \ref{fig:waterfall_a0_scan} as a function of the cascade growth rate. The data are well described by the linear relation
\begin{equation}
v/c = 0.45 \frac{\Gamma}{\omega_0},
\end{equation}
giving the logarithm a value of $\ln(a_{0}n_{c}/n_{s}) \approx 3.5$ within the expected range. This linear scaling with the cascade growth rate provides a direct and perhaps experimentally accessible link between the microscopic pair creation dynamics and the macroscopic plasma morphology observable in upcoming multi-petawatt experiments. With the spatial dynamics of the cascade now fully characterized, we turn in the last section to the question of how the total laser energy is partitioned between radiation, kinetic energy, and escaping light.

\section{Laser Energy Depletion}

The spatiotemporal dynamics raise a natural quantitative question: given the cascade growth dynamics and the temporal structure of realistic laser pulses, what fraction of the total laser energy is ultimately deposited into the plasma and radiation field? This was partially answered for linear polarization with a heuristic model \cite{Grismayer2016}. In this section, we verify whether the former findings also apply to circular polarization and extend the existing model to account for realistic time-dependent pulse envelopes.

\begin{figure}
\centering
\includegraphics[width=0.85\textwidth]{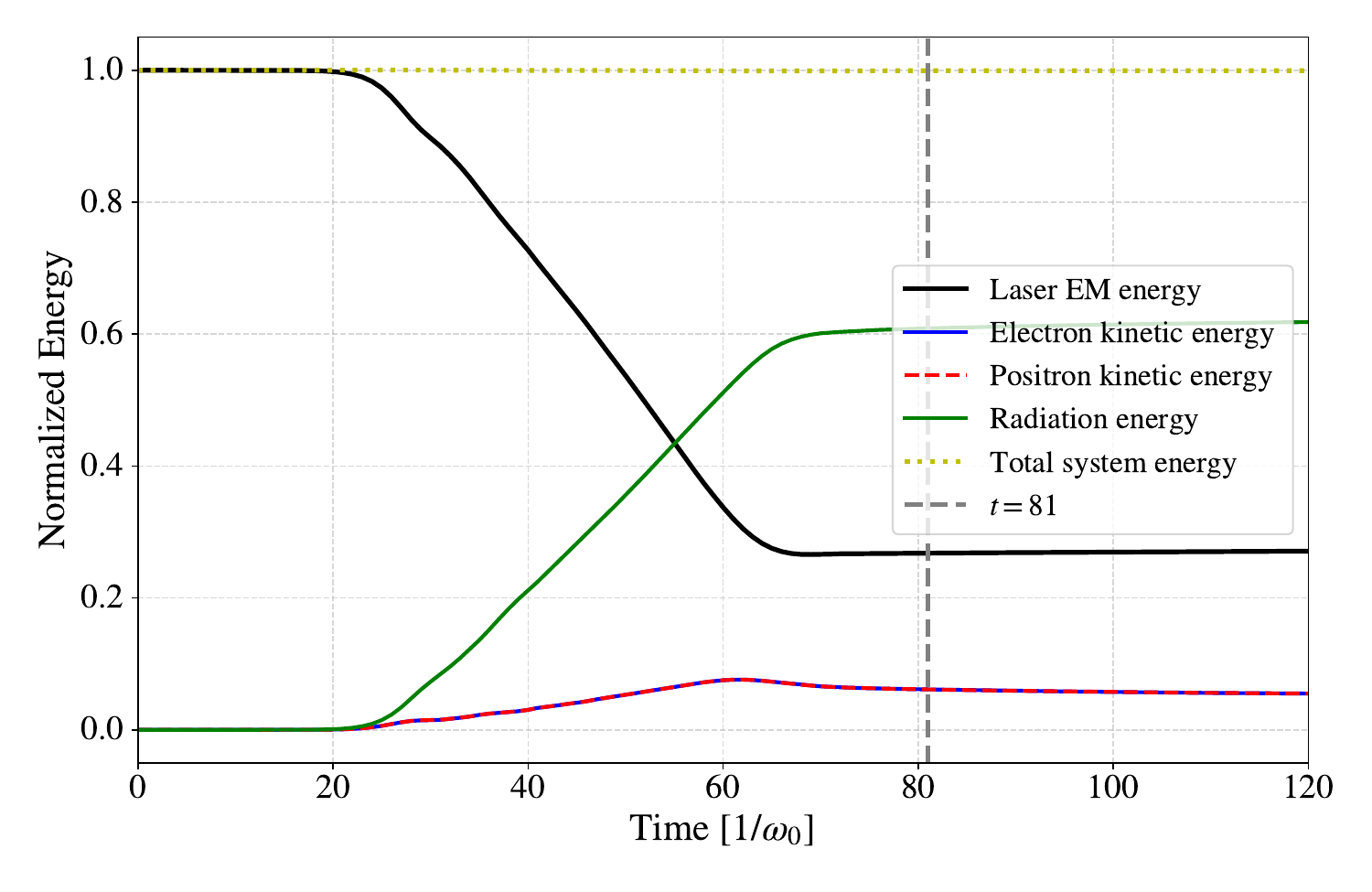}
\caption{Energy balance of a pair plasma generated by a self-sustained QED cascade in the rotating-field configuration for a laser amplitude of $a_{0} = 800$. The plot shows the depletion of the laser energy and its transfer primarily to emitted photons and the kinetic energy of the electron–positron pairs, resulting in approximately 70\% laser energy absorption. The line $t=81$ indicates the end of the laser pulse duration.}
\label{fig:energy partition}
\end{figure}

Figure \ref{fig:energy partition} illustrates the energy partition for $a_0 = 800$. The laser energy decreases monotonically from the onset of cascade-induced opacity, with the majority of the absorbed energy channeled into gamma-ray radiation rather than into the kinetic energy of the electron-positron pairs. This predominance of radiative energy conversion is consistent with previous findings in QED cascade and synchrotron-dominated plasma regimes \cite{Nerush2011_2, Bradly2012, Grismayer2017, Gonoskov2017, Vranic2019, Serebryakov2023}, and reflects the fact that nonlinear Compton scattering is far more frequent than pair creation at these field strengths. Only a small fraction of the absorbed energy is retained as particle kinetic energy, with the remainder escaping as unabsorbed laser light at the plasma boundaries. For $a_0 = 800$, the total laser absorption efficiency reaches approximately 70\%, which similar the level observed in linear polarization. 

To predict this absorption efficiency analytically, we build on the model \cite{Grismayer2016}, which assumed a flat-top temporal envelope of duration $\tau_0$ and amplitude $a_0$. In that framework, the absorption time, defined as the instant at which the plasma first reaches the relativistic critical density, determines the conversion efficiency 
\begin{equation}
\eta = \frac{\mathcal{E}_{a}}{\mathcal{E}_{T}} = 1-\frac{t_{a}}{\tau_{0}},
\end{equation}
where $\mathcal{E}_{T}$ and $\mathcal{E}_{a}$ are the total and absorbed laser energies respectively. While physically transparent, this ideal model is limited by its assumption of a perfectly flat temporal envelope, which overestimates the energy available during the intensity ramp and fall of realistic pulses.

To achieve a more realistic description, we incorporate a time-dependent normalized laser amplitude $a(t) = a_{0} f(t)$, where $f(t)$ defines the temporal envelope of the laser pulses. Because the cascade growth rate depends on the instantaneous laser amplitude, it becomes a time-varying quantity $\Gamma(a(t))$, and the condition for reaching the relativistic critical density must be generalized to
\begin{equation}
n_{0}\exp\left(\int_{0}^{t_{a}} \Gamma(t) dt\right) = a_{0}n_{c}.
\end{equation}
This implicitly defines the absorption time $t_a$ for an arbitrary envelope $f(t)$, after which the conversion efficiency takes the form
\begin{equation}
\eta = \frac{\int_{t_{a}}^{\tau_{0}}a^{2}(t)dt}{\int_{0}^{\tau_{0}}a^{2}(t)dt} = \frac{\int_{t_{a}}^{\tau_{0}}f^{2}(t)dt}{\int_{0}^{\tau_{0}}f^{2}(t)dt}.
\end{equation}
This generalization is valid under an adiabatic approximation, which requires that the cascade evolves on timescales much shorter than the characteristic variation time of the envelope, $\tau_{\text{env}}(t) = \vert f(t)/f'(t) \vert$, so that $1/\Gamma \ll \tau_{\text{env}}(t)$. In the high-field regime $a_0 > 600$, the cascade e-folding time is of order $1/\Gamma \sim \mathcal{O}(1\text{ fs})$ \cite{Grismayer2016, Mercuri2025}, while realistic intense laser pulses vary on timescales $\tau_{\text{env}} \sim \mathcal{O}(10-100\text{ fs})$ \cite{laser_power_in_time,eli, apolon}, ensuring a clear separation of timescales precisely during the active cascading window where the vast majority of pairs are generated. Note that assuming an ideal flat-top envelope, $f(t) = 1$, recovers the original conversion efficiency.

\begin{figure}
\centering\includegraphics[width=0.85\textwidth]{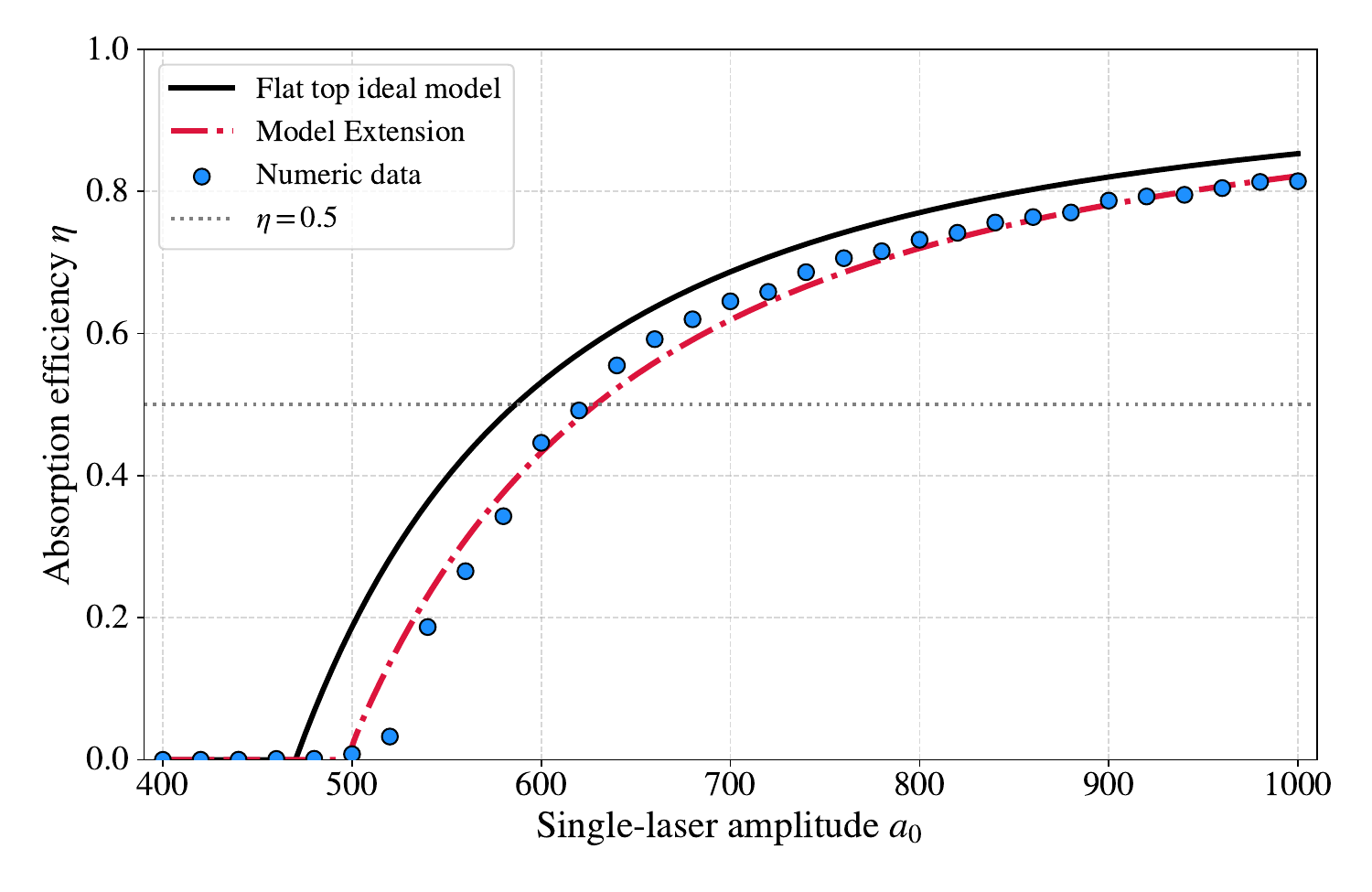}
\caption{Absorption efficiency, $\eta$, of a pair plasma generated by a self-sustained QED cascade in the rotating-field configuration across the parameter scan. The plot compares the numerical results for the absorption efficiency (blue dots) with the predictions of the ideal model (black line) and the extended model (red line). All datasets exhibit an asymptotic trend toward saturation.}
\label{fig: efficiency_results}
\end{figure}

Figure \ref{fig: efficiency_results} presents the absorption efficiency across the full parameter scan, comparing the numerical results with the predictions of both the ideal flat-top model and the extended model. The extended model accurately reproduces the numerical data across the entire intensity range explored, capturing in particular the shift in the onset amplitude at which significant laser depletion begins. Both models exhibit the same asymptotic trend toward complete laser energy conversion at high amplitudes, consistent with the picture of an increasingly efficient cascade as $a_0$ grows. The good agreement between the extended model and the simulations across the full parameter scan confirms that the adiabatic approximation provides an accurate and computationally inexpensive framework for predicting laser absorption in realistic experimental configurations. The absorption efficiency, the gamma-ray conversion fraction, and the onset amplitude of significant laser depletion constitute a set of robust and experimentally accessible macroscopic diagnostics for QED cascade formation.

\section{Conclusions}
We have investigated the spatiotemporal lifecycle of seeded QED cascades in a circularly polarized standing wave, from their initial threshold and exponential growth through radiative trapping, relativistic opacity, field disruption, and macroscopic laser energy depletion. Using 1D3V PIC-QED simulations with the code OSIRIS across a parametric scan of single-laser amplitudes $a_0 \in [300, 1000]$, we have established the physical picture of how these cascades saturate, how the plasma they generate disrupts the driving fields, and how the laser energy is ultimately partitioned between radiation, pairs, and escaping light.

The central result is that cascade saturation is governed by the competition between the growth rate and longitudinal lepton migration out of the magnetic nodes. Below a threshold amplitude, migration suppresses cascade growth, confining an under-dense plasma within the standing-wave pattern dictated by Normal Radiative Trapping at the electric nodes. Above this threshold, pair creation overwhelms migration, and the cascade drives the plasma density well beyond the relativistic critical density. Once the over-dense plasma is established, it reflects the incident laser pulses and disrupts the standing wave, replacing the stable periodic node structure with a dynamic interference pattern. This shifting pattern triggers secondary QED cascades in adjacent nodes, driving an apparent spatial expansion of the critical density surface. 

We have shown that the vast majority of absorbed laser energy is channeled into gamma-ray radiation rather than particle kinetic energy, consistent with the synchrotron-dominated nature of the interaction. By extending the model \cite{Grismayer2016} to incorporate realistic time-dependent pulse envelopes under an adiabatic approximation, we have derived a generalized conversion efficiency that accurately reproduces the simulated absorption across the full intensity range explored. 

These findings have direct and concrete implications for the design and interpretation of experiments at upcoming multi-petawatt laser facilities. Three robust macroscopic diagnostics of over-dense pair plasma formation emerge from this work: the onset of wave reflection, evidenced by a sign reversal of the Poynting flux at the plasma boundary; the apparent plasma expansion velocity; and the fraction of laser energy converted into gamma-ray photons, which provides a signature of cascade activity accessible to existing gamma-ray detectors. Together, these observables offer a coherent and experimentally accessible framework for verifying QED cascade formation in the laboratory without requiring direct access to the microscopic pair creation dynamics.

While the 1D geometry adopted here captures the fundamental longitudinal dynamics and establishes an upper bound on laser absorption, it necessarily neglects transverse effects that will be present in realistic focused laser beams. In particular, ponderomotive expulsion of plasma from the focal volume and finite focal spot geometries are expected to modify both the lepton migration rate and the threshold for relativistic opacity. Extending the present analysis to two and three dimensions, and assessing how transverse dynamics alter the expansion threshold and the absorption efficiency, are therefore important directions for future work. Nevertheless, the physical mechanisms identified here (particle creation inertia, standing wave disruption, and expansion-driven secondary cascades) are expected to remain relevant in higher-dimensional geometries, and the analytical framework developed provides a natural starting point for such extensions.

\bibliographystyle{unsrt} 
\bibliography{bib/ref}


\end{document}